\documentclass{PoS}

\newcommand{\ovl}[1]{\overline{#1}}

\newcommand{\p}{\partial}

\newcommand{\pslash}{p\kern-1ex /}
\newcommand{\lslash}{l\kern-1ex /}
\newcommand{\kslash}{k\kern-1ex /}
\newcommand{\dslash}{\p\kern-1.2ex /}
\newcommand{\Dslash}{{\cal D}\kern-1.5ex /}
\newcommand{\Aslash}{A\kern-1.2ex /}

\newcommand{\vev}[1]{\left\langle #1 \right\rangle}

\newcommand{\les}{\stackrel{<}{{}_{\sim}}}

\def\tfrac#1#2{{\textstyle\frac{#1}{#2}}}

\title{Light meson spectrum with $N_f=2+1$ dynamical overlap fermions }

\ShortTitle{Light meson spectrum with $N_f=2+1$ dynamical overlap fermions }

\author{
  JLQCD and TWQCD collaborations:
  \speaker{J.~Noaki}$^{,a}$\thanks{E-mail: noaki@post.kek.jp},
  S.~Aoki$^{b,c}$, 
  T.W.~Chiu$^d$,
  H.~Fukaya$^{a,e}$,
  S.~Hashimoto$^{a,f}$,
  T.H.~Hsieh$^g$,
  T.~Kaneko$^{a,f}$, 
  H.~Matsufuru$^a$,
  T.~Onogi$^h$,
  E.~Shintani$^a$ and
  N.~Yamada$^{a,f}$
  \vspace*{2mm}
  \\
  \llap{$^a$}
  High Energy Accelerator Research Organization (KEK),
  Tsukuba 305-0801, Japan
  \\
  \llap{$^b$}
  Graduate School of Pure and Applied Sciences,
  University of Tsukuba, Tsukuba 305-8571, Japan
  \\
  \llap{$^c$}
  Riken BNL Research Center, Brookhaven National Laboratory, Upton, NY
  11973, USA
  \\
  \llap{$^d$}
  Physics Department, Center for Theoretical Sciences,
  and Center for Quantum Science and Engineering,
  National Taiwan University, Taipei 10617, Taiwan
  \\
  \llap{$^e$}
  The Niels Bohr Institute, The Niels Bohr International Academy,
  Blegdamsvej 17 DK-2100 Copenhagen {\O}, Denmark
  \\
  \llap{$^f$}
  School of High Energy Accelerator Science,
  the Graduate University for Advanced Studies (Sokendai),
  Tsukuba 305-0801, Japan
  \\
  \llap{$^g$}
  Research Center for Applied Sciences, Academia Sinica,
  Taipei~115, Taiwan
  \\
  \llap{$^h$}
  Yukawa Institute for Theoretical Physics, 
  Kyoto University, Kyoto 606-8502, Japan
}

\abstract{
   We report on a numerical simulation with 2+1 dynamical flavors of 
   overlap fermions. We calculate pseudo-scalar masses and decay
   constants on a $16^3\times 48 \times (0.11\ {\rm fm})^4$ lattice 
   at five different up and down quark masses and two strange quark masses.
   The lightest pion mass corresponds to $\approx 310$ MeV.
   We also study the validity of the chiral perturbation theory 
   using the results of the numerical simulation with two dynamical flavors 
   and conclude that the one-loop formulae cannot be directly applied in
   the strange quark mass region. 
   We therefore extrapolate our 2+1-flavor results to 
   the chiral limit by fitting the data to the two-loop formulae of the
   chiral perturbation theory.
}
   
\FullConference{The XXVI International Symposium on Lattice Field Theory\\
		 July 14-19 2008\\
		 Williamsburg, Virginia, USA}

\begin{document}

\section{Introduction}

The lattice simulation with the overlap fermions~\cite{Neuberger1998}
provides a theoretically ideal setup to study low energy hadron physics.
With the exact chiral symmetry, the continuum chiral perturbation theory
(ChPT) can be applied without modification in sharpe contrast to other 
fermion formulations that violate either chiral or flavor symmetry.
For this advantage, we performed numerical simulation using
overlap fermion action with two dynamical flavors~\cite{Nf2_generation}.
In particular, we carried out the calculation of light meson spectrum
and studied the consistency between QCD and chiral perturbation theory.
In this article, we present an extension of this study to the $N_f=2+1$
QCD similar parameters. 
On a $16^3\times 48$ lattice, we generate 2,500
trajectories~\cite{HashimotoProc,MatsufuruProc} at ten different
combinations of up/down and strange sea quark masses, {\it i.e.} 
five $m_{ud}$'s and two $m_s$'s. As in the $N_f=2$ case, 
topological charge of the gauge configuration is fixed to zero throughout 
Monte Carlo updates.
Lattice spacing is determined by the Sommer scale as $a=0.1075(7)$ fm 
or $a^{-1}=1.833(12)$ GeV from an input $r_0=0.49$ fm. 

In section~\ref{spect_section}, we present basic part of the calculation 
of pseudo-meson masses and decay constants. 
Since these results depend on strange quark mass, the chiral
extrapolation should be performed with a fit ansatz valid beyond
the scale of kaon mass. To discuss this issue, we review the test of ChPT 
performed for the $N_f=2$ case in section~\ref{Nf2chiral}. 
Based on this test, in section~\ref{Nf2+1chiral}, we present 
the extrapolation of the $N_f=2+1$ data by using the two-loop ChPT formulae.

\section{Spectrum calculation ($N_f=2+1$)}\label{spect_section}

\begin{figure}
 \begin{center}
  \includegraphics[width=7.6cm,clip]{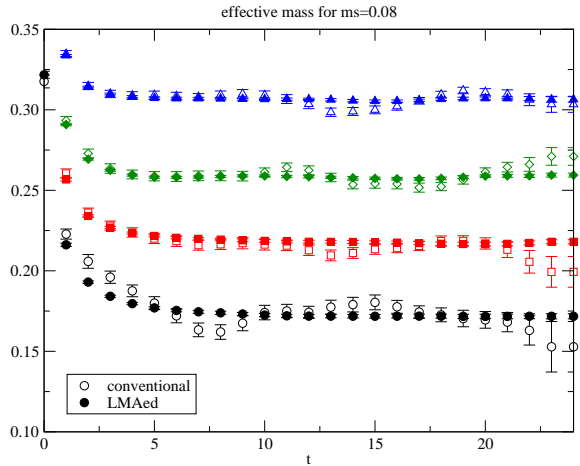}
  \caption{Effective pion mass plots for the four lightest quark mass.
  Open symbols indicate signals from conventional treatment while filled
  symbols are from low-mode-averaging.}
  \label{Meff}
 \end{center}
\end{figure}

For the $N_f=2+1$ case, we calculate 80 pairs of the lowest-lying 
eigenmodes on each gauge configuration and store them on the disks. 
These eigenmodes are used to construct the low-mode contribution to 
the quark propagators. The higher-mode contribution is obtained by 
conventional CG calculation with significantly smaller amount of
machine time than the full CG calculation. Those eigenmodes are also used to 
replace the lower-mode contribution in the meson correlation functions
by that averaged over the source location 
(low-mode averaging)~\cite{DeGrand2004,Giusti2004}. 
The noise of correlation function is decreased as shown in
Figure~\ref{Meff}, where effective pion masses are compared between
the data with and without the low-mode averaging for the lightest four quark 
masses.

In order to obtain renormalization factor of quark mass $Z_m$, we
calculate scalar and pseudo-scalar vertex functions in the momentum 
space in the Landau gauge and applied the RI/MOM scheme~\cite{Martinelli1995}.
In the fit of the vertex functions we explicitly use the low-mode contribution
to the chiral condensate
\begin{eqnarray}
 \vev{\bar{q}q}(m_q) = 
  \vev{\frac{1}{V}\sum_{i=1}^{80}\frac{2m_q}{m_q^2+\lambda_i^2}}
\end{eqnarray}
with eigenvalues $\lambda_i$ and absorb its valence quark mass
dependence. By converting $Z_m^{\rm RI/MOM}$
in the massless limit into the value of $\ovl{\rm MS}$ perturbatively, 
we obtain a preliminary value 
$Z_m^{\ovl{\rm MS}}(2\,{\rm GeV})=0.815(8)$.

On our lattice with spatial extent $L$, the lightest pion gives 
$m_\pi L\approx 2.8$. Finite size effect (FSE) may therefore be sizable. 
We estimate this effect using the analytic result
~\cite{Colangelo2005} from a combination of the resummed L\"uscher's
formulae and an one-loop ChPT. There is an additional finite size 
due to the fixed topology in our numerical simulation~\cite{Brower2003}. 
Based on the discussion in~\cite{FSE_Chit}, we make 
a correction using the result of one-loop ChPT and the numerical data of 
topological susceptibility~\cite{Chiu_writeup} determined on the same 
lattice configurations.

\section{Chiral extrapolation ($N_f=2$)}\label{Nf2chiral}

We address the validity of the NLO ChPT prediction which is commonly 
used in the chiral extrapolation of spectrum data on the lattice
~\cite{Nf2_spectrum}.
In the framework of $N_f=2$, pion mass and decay constants are 
expanded in terms of $x= 4Bm_q/(4\pi f)^2$ as
\begin{eqnarray}
 m_\pi^2/m_q &=& 2B(1+\tfrac{1}{2}x\ln x) +c_3 x,\\
 f_\pi       &=& f(1 -x\ln x) +c_4 x
\end{eqnarray}
to NLO ({\it i.e.} one-loop level or ${\cal O}(x)$), where 
$B$ and $f$ are the tree level low energy constants (LECs), and 
$c_3$ and $c_4$ are related to the one-loop level LECs $\bar{l}_3$ and 
$\bar{l}_4$.
At NLO, these expressions are unchanged when one replaces the 
expansion parameter $x$ by
$\hat{x}=2m_\pi^2/(4\pi f)^2$ or $\xi=2m_\pi^2/(4\pi f_\pi)^2$, 
where $m_\pi^2$ and $f_\pi$ denote those at a finite quark mass.
Therefore, in a small enough pion mass region the three expansion 
parameters should describe the lattice data equally well.

\begin{figure}
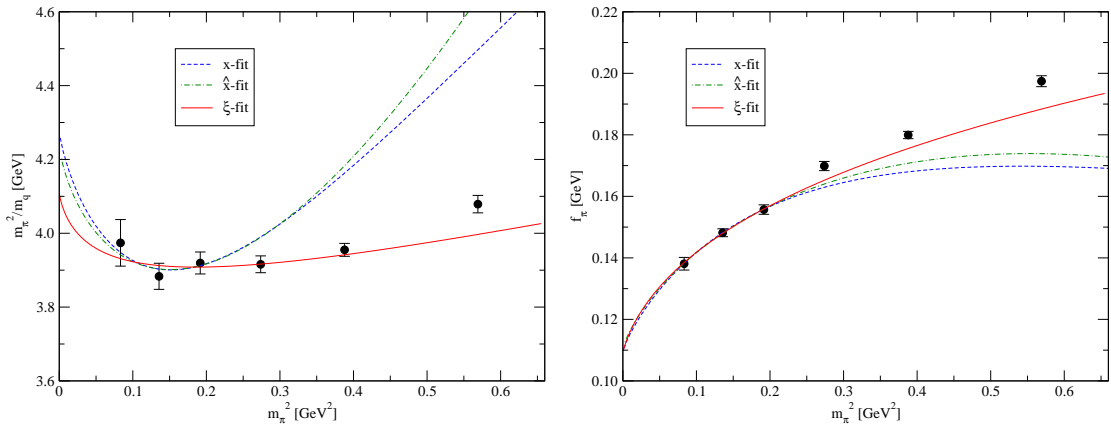

 \begin{center}
  \includegraphics[width=7.2cm,clip]{mp2r_nlo.eps}
  \hspace{1mm}
  \includegraphics[width=7.2cm,clip]{fps_nlo.eps}
  \caption{Comparison of the chiral fits including the NLO terms for 
  $m_\pi^2/m_q$ (left) and $f_\pi$ (right) obtained in the $N_f=2$ 
  calculation. Fit curves for the three lightest
  data points obtained with different choices of the expansion parameter
  ($x$, $\hat{x}$ and $\xi$) are shown as a function of $m_\pi^2$.}
  \label{Nf2_NLOchpt}
 \end{center}
\end{figure}

Three fit curves ($x$-fit, $\hat{x}$-fit and $\xi$-fit) for the 
three lightest pion mass points ($m_\pi\les 450$ MeV) are shown in 
Figure~\ref{Nf2_NLOchpt} as a function of $m_\pi^2$. 
For all fits, the horizontal axis is appropriately rescaled to give $m_\pi^2$
using the obtained fit curves.

\begin{figure}
 \begin{center}
  \includegraphics[width=7.8cm,clip]{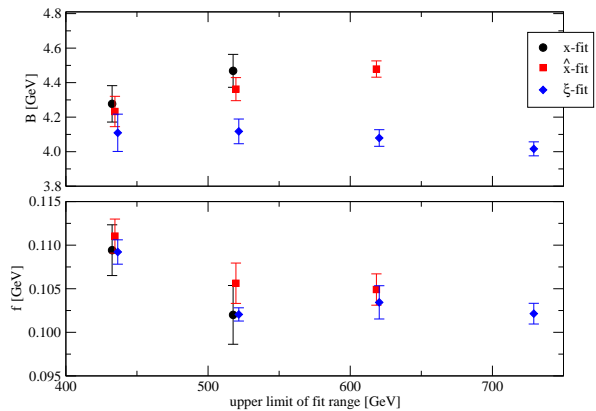}
  \caption{Results of fit parameters $B$ (top) and $f$ (bottom) 
  as functions of the upper limit of the fit range. In each panel,
  circle, square and diamonds are obtained with fit parameters $x$,
  $\hat{x}$ and $\xi$. Results with $\chi^2/$dof $\les 2$ are plotted. }
  \label{Nf2coeff}
 \end{center}
\end{figure}

From the plot we 
observe that the different expansion parameters describe the 
three lightest points equally well; the values of $\chi^2/$dof are 0.30, 
0.33 and 0.66 for $x$-, $\hat{x}$- and $\xi$-fits, respectively. 
In each fit, the correlation between 
$m_\pi^2/m_q$ and $f_\pi$ for common sea quark mass is taken into account.
Between the $x$- and $\hat{x}$-fit, all of the resulting fit parameters are 
consistent. Among them, $B$ and $f$ are also consistent with the $\xi$-fit.
This indicates that the NLO formulae successfully describes the data. 
In Figure~\ref{Nf2coeff}, results of $B$ (upper panel) and $f$
(bottom panel) for different fits are plotted for different pion 
mass points. As seen in the figure, the agreement among the different 
expansion prescriptions is lost when we extend the fit range to include 
the 4th lightest data point which corresponds to $m_\pi\simeq$ 520~MeV.
We, therefore, conclude that for these quantities the NLO ChPT may be 
safely applied only below $\approx$ 450~MeV.

Another important observation from Figure~\ref{Nf2_NLOchpt} is that only
the $\xi$-fit reasonably describes the data beyond the fitted region.
With the $x$- and $\hat{x}$-fits the curvature due to the chiral
logarithm is too strong to accommodate the heavier data points.
In fact, values of the LECs with the $x$- and $\hat{x}$-fits 
are more sensitive to the fit range than the $\xi$-fit.
This is because $f$, which is significantly smaller than $f_\pi$ of our 
data, enters in the definition of the expansion parameter.
Qualitatively, by replacing $m_q$ and $f$ by $m_\pi^2$ and $f_\pi$ 
the higher loop effects in ChPT are effectively resummed and the
convergence of the chiral expansion is improved.

\begin{figure}
 \begin{center}
  \includegraphics[width=7.8cm,clip]{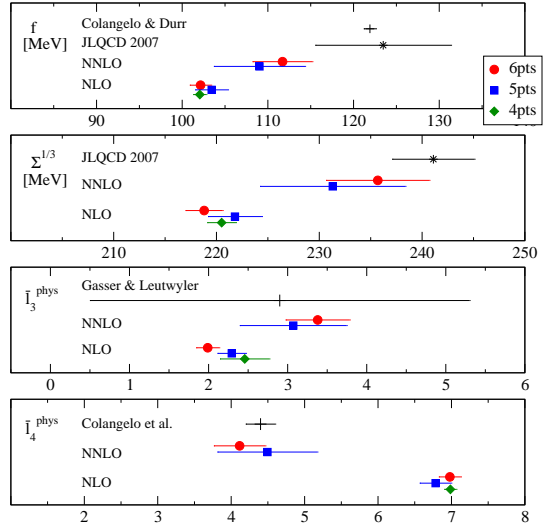}
  \caption{Comparison of the $N_f=2$ results from the NLO fit and 
  the NNLO fit with $\xi$. Black pluses denote reference points
  from phenomenological estimations.}
  \label{all_pion_cmpr}
 \end{center}
\end{figure}

We then extend the analysis to include the NNLO terms~\cite{Colangelo2001}.
Since we found that only the $\xi$-fit reasonably describes the data
beyond $m_\pi\simeq$ 450~MeV, we perform the NNLO analysis using the
$\xi$-expansion. 
Although we input phenomenological estimate of LECs
$\bar{l}_1$ and $\bar{l}_2$, we find our fit result is insensitive to
their uncertainties.
We extract the LECs of ChPT, {\it i.e.} 
the decay constant in the chiral limit $f$, chiral condensate
$\Sigma= Bf^2/2$, and the NLO LECs
$\bar{l}_3^{\rm phys}= -c_3/B +\ln (2\sqrt{2}\pi f/m_{\pi^+})^2$ and 
$\bar{l}_4^{\rm phys}= c_4/f +\ln (2\sqrt{2}\pi f/m_{\pi^+})^2$.
For each quantity, a comparison of the results between the NLO and the 
NNLO fits is shown in Figure~\ref{all_pion_cmpr}. 
In each panel, the results with 5 and 6 lightest data points are plotted
for the NNLO fit. 
The correlated fits give $\chi^2/$dof = 1.94 and 1.40, respectively. 
For the NLO fits, we plot results obtained with 4, 5 and 6 points to
show the stability of the fit. The $\chi^2/$dof is less than 1.94.
The results for these physical quantities are consistent within either
the NLO or the NNLO fit. On the other hand, as seen for 
$\bar{l}_4^{\rm phys}$ most prominently, there is a significant 
disagreement between NLO and NNLO. 
This is due to the large NNLO contributions to the terms 
which are proportional to $c_3$ and $c_4$, respectively.

We quote our final results for the $N_f=2$ calculation from the NNLO 
fit with all data points:
$f=111.7(3.5)(1.0)(^{+6.0}_{-0.0})$~MeV,
$\Sigma^{\ovl{\rm MS}}(\mathrm{2~GeV})=
 [235.7(5.0)(2.0)(^{+12.7}_{-\ 0.0})\mathrm{~MeV}]^3$, 
$\bar{l}_3^{\rm phys}=3.38(40)(24)(^{+31}_{-\ 0})$, and
$\bar{l}_4^{\rm phys}=4.12(35)(30)(^{+31}_{-\ 0})$, where
$m_\pi^+=139.6$ MeV.
From the value at the neutral pion mass $m_{\pi^0}=135.0$ MeV, 
we obtain the average up and down quark mass $m_{ud}$ and the pion
decay constant as 
$m_{ud}^{\ovl{\rm MS}}(\mathrm{2~GeV})
=4.452(81)(38)(^{+\ 0}_{-227})$~MeV
and $f_\pi =119.6(3.0)(1.0)(^{+6.4}_{-0.0})$~MeV.
In these results, the first error is statistical, 
where the error of the renormalization constant is included in quadrature 
for $\Sigma^{1/3}$ and $m_{ud}$.
The second error is systematic due to the truncation of the higher
order corrections.
For quantities carrying mass dimensions, the third error is from 
the ambiguity in the determination of $r_0$. 
We estimate these errors from the difference of the results with our
input $r_0=0.49$~fm and that with $0.465$~fm.
The third errors for $\bar{l}_3^{\rm phys}$ and $\bar{l}_4^{\rm phys}$
reflect an ambiguity of choosing the renormalization scale of ChPT
($4\pi f$ or $4\pi f_\pi$). 

\section{Chiral extrapolation ($N_f=2+1$)}\label{Nf2+1chiral}

\begin{figure}
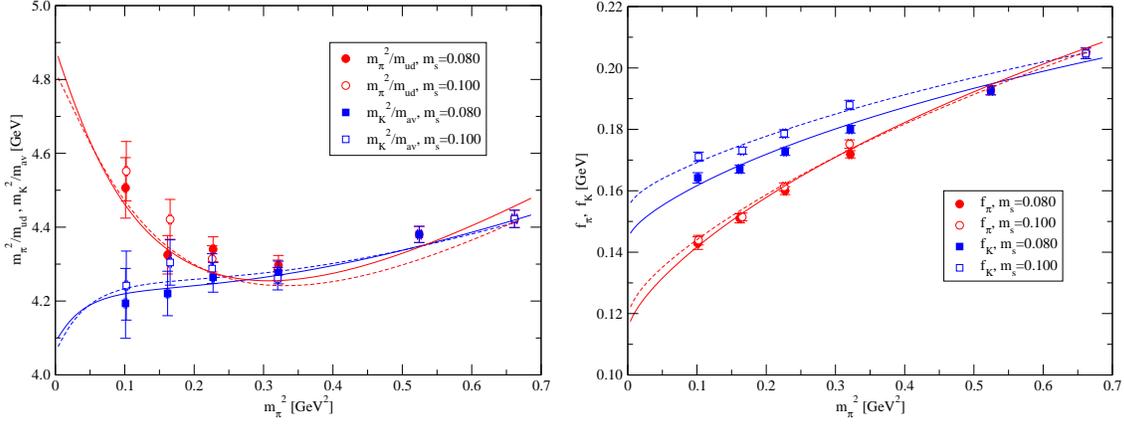

 \begin{center}
  \includegraphics[width=7.3cm,clip]{mpmk2r_nnlo.eps}
  \hspace{1mm}
  \includegraphics[width=7.3cm,clip]{fpfk_nnlo.eps}
  \caption{Chiral properties of $m_\pi^2/m_{ud}$ and 
  $m_K^2/m_{av}$ (left) and $f_\pi$ and $f_K$ (right) for fixed 
  values of strange quark mass
  (open and filled symbols). In each panel, circles (squares) indicate 
  the pion (kaon) data. }
  \label{Nf2+1chpt}
 \end{center}
\end{figure}

Since we found in the two-flavor calculation that the NNLO ChPT formulae
can nicely fit our data even in the kaon mass region 
if one uses the $\xi$-expansion,
we apply the same strategy for our $2+1$-flavor analysis.

As functions of $\xi_\pi=m_\pi^2/(4\pi f_\pi)^2$ and 
$\xi_{\eta_s}=m_{\eta_s}^2/(4\pi f_\pi)^2$, where $\eta_s$ is the unphysical 
strange-strange meson, chiral expansions are expressed as 
\begin{eqnarray}
 m_\pi^2/m_{ud} &=& 2B_0(1+ M^\pi(\xi_\pi, \xi_{\eta_s}; L_i^r)) 
                    +\alpha_1^\pi\xi_\pi^2 
                    +\alpha_2^\pi\xi_\pi\xi_{\eta_s} 
                    +\alpha_3^\pi\xi_{\eta_s}^2 \label{mp2rSU3}\\
 m_K^2/m_{av} &=& 2B_0(1+ M^K(\xi_\pi, \xi_{\eta_s}; L_i^r))  
                  +\alpha_1^K\xi_\pi(\xi_\pi -\xi_{\eta_s})
      	          +\alpha_2^K\xi_{\eta_s}(\xi_{\eta_s} -\xi_\pi)
		  \label{mk2rSU3}\\
 f_\pi &=& f_0(1+ F^\pi(\xi_\pi, \xi_{\eta_s}; L_i^r)) 
           +\beta_1^\pi\xi_\pi^2 
           +\beta_2^\pi\xi_\pi\xi_{\eta_s} 
           +\beta_3^\pi\xi_{\eta_s}^2\label{fpiSU3}\\
 f_K &=& f_0(1+ F^K(\xi_\pi, \xi_{\eta_s}; L_i^r))  
          +\beta_1^K\xi_\pi(\xi_\pi -\xi_{\eta_s})
	  +\beta_2^K\xi_{\eta_s}(\xi_{\eta_s} -\xi_\pi),\label{fKSU3}
\end{eqnarray}
where $m_{av} = \tfrac{1}{2}(m_s +m_{ud})$ and $\alpha^{\pi,K}_i$
and $\beta^{\pi,K}_i$ are NNLO unknown parameters.
Functions $M^\pi$, $M^K$, $F^\pi$ and $F^K$ contain NLO contributions
and loop contributions at NNLO, whose expressions are too involved to 
present here~\cite{Amorosetal2000}. 
Among relevant SU(3) LECs $L_1^r$--$L_8^r$, we use values
$L_1^r =(0.38\pm 0.18)\cdot 10^{-3}$, 
$L_2^r =(1.59\pm 0.15)\cdot 10^{-3}$, 
$L_3^r =(-2.91\pm 0.32)\cdot 10^{-3}$ and 
$L_7^r =(-0.49\pm 0.24)\cdot 10^{-3}$ (defined at $\mu=770$ MeV) 
from a phenomenological estimate~\cite{Amorosetal2001} and determine
others by a fit. Thus, the chiral extrapolation
with (\ref{mp2rSU3})--(\ref{fKSU3}) contains 16 fit parameters in total.
We fit $m_\pi^2/m_{ud}$, $m_K^2/m_{av}$, $f_\pi$ and $f_K$
simultaneously taking the correlation between these quantities
at the same sea quark mass $(m_{ud}, m_s)$ into account. By using 
all data points, $\chi^2/$dof = 1.42 is obtained.
Data after the finite size corrections are shown in 
Figure~\ref{Nf2+1chpt} as a function of $m_\pi^2$. 
Different symbols correspond to the pion data 
($m_\pi^2/m_{ud}$ and $f_\pi$) and the kaon data ($m_K/m_{av}$ and
$f_K$) while the filled (open) pattern represent a
fixed lighter (heavier) strange quark mass, which is accompanied by 
the solid (dashed) curves.

Extrapolating the data to the physical point $m_{\pi}=135.0$ MeV, 
$m_K=495.0$ MeV and $f_\pi=130.7$ MeV, we obtain preliminary results
$m_{ud}^{\ovl{\rm MS}}(2\ {\rm GeV}) = 3.799(68)\ {\rm MeV}$,
$m_s^{\ovl{\rm MS}}(2\ {\rm GeV}) = 114.6(2.0)\ {\rm MeV}$,
$f_\pi = 121.5(4.1)\ {\rm MeV}$,
$f_K   = 148.3(4.7)\ {\rm MeV}$ and
$f_K/f_\pi = 1.220(10)$,
where the errors are statistical only. 
Further studies to determine LECs and to analyze their flavor dependence
are planned as well as the estimation of systematic errors.

\vspace*{6mm}
Numerical simulations are performed on Hitachi SR11000 and
IBM System Blue Gene Solution at High Energy Accelerator Research
Organization (KEK) under a support of its Large Scale
Simulation Program (No.~07-16). 
This work is supported in part by the Grant-in-Aid of the
Ministry of Education 
(Nos.
18340075,
18740167,
19540286,
19740121,
19740160,
20025010,
20039005,
20340047,
20740156),
the National Science Council of Taiwan 
(Nos. NSC96-2112-M-002-020-MY3,
      NSC96-2112-M-001-017-MY3,
      NSC97-2119-M-002-001)
and NTU-CQSE (No. 97R0066-69).
\newcommand{\J}[4]{{#1} {\bf #2} (#3) #4}
\newcommand{\RMP}{Rev.~Mod.~Phys.}
\newcommand{\MPL}{Mod.~Phys.~Lett.}
\newcommand{\IJMP}{Int.~J.~Mod.~Phys.}
\newcommand{\NP}{Nucl.~Phys.}
\newcommand{\NPSup}{Nucl.~Phys.~{\bf B} (Proc.~Suppl.)}
\newcommand{\PL}{Phys.~Lett.}
\newcommand{\PRD}{Phys.~Rev.~D}
\newcommand{\PRL}{Phys.~Rev.~Lett.}
\newcommand{\AP}{Ann.~Phys.}
\newcommand{\CMP}{Commun.~Math.~Phys.}
\newcommand{\CPC}{Comp.~Phys.~Comm.}
\newcommand{\PTP}{Prog. Theor. Phys.}
\newcommand{\Suppl}{Prog. Theor. Phys. Suppl.}
\newcommand{\JHEP}{JHEP}

\end{document}